\documentclass[10pt]{article}

\def\@cite#1#2{$^\mbox{{\footnotesize #1\if@tempswa,#2\fi}}$}

\begin{document}

\begin{center}
{\large \bf New coorbital dynamics in the solar system}\\[2mm]
 F. Namouni, A. A. Christou and C. D. Murray\\[2mm]
{\small \it Astronomy Unit, Queen Mary and Westfield College,\\ 
Mile End Road, London E1 4NS, United Kingdom}\\[5mm]

{\bf Abstract}\\ \end{center}
Following the discovery that asteroid (3753) Cruithne was a coorbital
companion of the Earth, a new theory of coorbital motion has been developed 
whereby 
planets or satellites can maintain companion objects in the same
orbit as themselves. This has led to the prediction of
hitherto unknown types of stable motion, all of which are seen in 
the evolution of specific near-Earth asteroids.  
The slow diffusion of such objects through the Earth's coorbital region  
is shown to lead to temporary capture,  suggesting the existence of 
undiscovered retrograde moons of the Earth.

\vspace*{5mm} 
\begin{center} {\bf 1. Introduction}\\[2mm] \end{center}  
 Lagrange$^1$ demonstrated the existence of five equilibrium
positions in the three-body problem. 
Those that form an equilateral triangle with the Sun and the planet, 
the leading $L_4$ point and trailing $L_5$ point,  are stable to
small displacements.  However, it was not until the subsequent
discovery of the first Trojan asteroid, (588) Achilles, near the
$L_4$ point in the Sun-Jupiter system that a real example of such
motion was first observed.  There are now known to be at least 400
Trojan asteroids in the orbit of Jupiter, and at least one, (5261)
Eureka, near Mars's
$L_5$ point;$^2$ there are also three examples of Trojan moons in
the Saturnian system.$^3$  In fact, all the
known Trojan objects are librating about $L_4$ or
$L_5$ points and their orbits are often referred to as {\it
tadpoles\/} (or T orbits) because of their shape with respect to
the equilibrium position, as viewed in the frame rotating with the mean angular
velocity (mean motion) of the planet (see Fig~1).  
 
When the angular amplitudes of the leading and trailing 
tadpoles are sufficiently large, the orbits  
merge near the unstable $L_3$ point located
$180^\circ$ from the planet; the resulting paths in the rotating
frame are referred to as {\it horseshoes\/} (or H orbits). A 
variation on such a structure is seen in the orbits of Janus and
Epimetheus, the coorbital satellites of Saturn.$^4$  In this
particular case, a good theoretical understanding of the peculiar
orbital dynamics is possible because of the low eccentricities
($e<0.01$) and inclinations ($I<0.4^\circ$) of both objects.  In
contrast, it has recently been discovered that asteroid (3753)
Cruithne, previously designated 1986TO, performs a {\it
temporary\/} horseshoe-like orbit with respect to the Earth.$^5$ 
This is a more extreme example that has been difficult to
incorporate in an analytical theory  because of Cruithne's large
eccentricity ($e=0.515$) and inclination ($I=19.8^\circ$).

In this article, we present aspects of the general theory of 
coorbital motion that accounts for all types of coorbital behavior.$^6$ 
This shows that asteroid (3753) Cruithne is only a member of a more general 
class of objects that are captured in the coorbital regions of the planets, 
and that involve new types of coorbital motion at large eccentricity
and inclination.  We also report the identification of these new orbits
in the motion of specific asteroids.
For the cases we have examined, the additional planetary
perturbations can cause near-Earth objects to be trapped into the 
coorbital regions of the terrestrial planets without any risk of collisions.
The structures we found provide a new paradigm for coorbital motion in 
the three-body problem and we believe that they have played an important 
role in such diverse problems as the origin of retrograde outer 
satellites and the accretional processes in the early solar system. 

\vspace*{5mm}

\begin{center} {\bf 2. Types of motion}\\[2mm] \end{center}
In the Sun-planet-asteroid three-body problem, the motion of an 
asteroid on a small eccentricity and inclination orbit is constrained by 
the value of the Jacobi constant through the
`excluded' regions associated with the zero velocity curves$^7$ (Fig~1). 
The T and H librations are found to be limited to the
annulus between the Lagrangian points $L_1$ and $L_2$ defined as  the
coorbital region of the planet; its half-width equals the radius of the
planet's sphere of influence: 
\begin{equation}
\Delta_{\rm P}=\left[m_{\rm P}/
3(m_{\rm P}+m_{\rm S})\right]^{1/3} a_{\rm P}
\end{equation} where $a_{\rm P}$ is the semi-major axis of the
planet's orbit, and $m_{\rm P}$ and $m_{\rm S}$ are the masses of the
planet and the Sun respectively. 
The outer boundary of the coorbital region is made up of irregular orbits 
and separates the H orbits from the passing orbits (or P orbits).  

For large eccentricities, 
the previous picture of coorbital motion is modified substantially 
because the Jacobi constant can no longer exclude spatial regions for 
the motion of the asteroid. In addition, the large eccentricity of the 
asteroid's orbit can lead to physical collisions with the planet. 
There are two facts that allows us to develop a theory that is valid
for all types of coorbital motion: the first is the realisation that
the asteroid's motion can be described as the superposition of two types
of motion: (i) a  fast three dimensional gyration with amplitudes of
$O(a_{\rm A}e)$ and $O(a_{\rm A}\sin I)$ and a frequency equal to 
the asteroid's mean
motion and (ii) the slow evolution of the mean position of the asteroid 
with respect to the planet. The latter is referred to as the motion of
the {\it guiding center} and is  represented by the
conjugate variables: relative semi-major axis, 
$a_{\rm r}=(a_{\rm A}-a_{\rm P})/ a_{\rm P}$, and relative mean longitude, 
$\lambda_{\rm r}=\lambda_{\rm A}-\lambda_{\rm P}$.
 
The second fact is the regularity of $e$, $I$  and the
argument of perihelion $\omega$ away from the boundary  of the coorbital
region; this allows us to derive the motion of the guiding center
by perturbation analysis. The Jacobi constant can then be used to  
constrain the motion of the guiding center instead of the full
motion, by averaging with respect to the 
fast gyration represented by $\lambda_{\rm A}$. 
The guiding center $a_{\rm r}(\lambda_{\rm r})$ can be written as:
\begin{equation}
a_{\rm r}^2=C-\frac{8m_{\rm P}}{3(m_{\rm P}+m_{\rm S})}\,
S(\lambda_{\rm r},e,I,\omega)
\end{equation}
where $C$ is a constant that defines the nature of the orbit, and $S$
is related to the averaged interaction potential by:$^6$
\begin{equation}    
S=\left. \frac{1}{2\pi}\int_{-\pi}^{\pi} \left( |{\bf r}_{\rm
A}-{\bf r}_{\rm P}|^{-1}-{\bf r}_{\rm A}\cdot {\bf r}_{\rm P}\right)
{\rm d}\lambda_{\rm A} \right|_{a_{\rm r}=0}.
\end{equation}
In equation (3), ${\bf r}_{\rm A}$ and ${\bf r}_{\rm P}$ are the
position vectors, and  the second term is due to the
motion of the Sun; the distances are scaled to $a_{\rm P}$. 
The motion along the guiding center can now be viewed as that of a
particle in the unidimensional potential well $S$ (Fig~2). 
 
The fact that the shape of the guiding center  depends
on  $e$,  $I$ and $\omega$, has three major consequences 
for the coorbital motion: firstly, the effective 
Lagrangian points $L_4$ and $L_5$ can be displaced appreciably from
the equilateral configuration (Fig~2 b,c,d). In addition, they  
no longer  
correspond necessarily to the same Jacobi constant, which fact implies an
asymmetry in the distribution of T orbits.$^8$

Secondly, retrograde satellite orbits (or RS orbits) appear outside the 
planet's sphere of influence as they are the only family of bounded orbits 
around the planet that remain stable with increasing
eccentricity$^9$ (Compare a  and b in Fig~2). 
For the planar motion, it can be seen that this family is separated 
from the H and T orbits by the collision singularity, 
${\bf r}_{\rm A}= {\bf r}_{\rm P}$,
at $\lambda_{\rm r}\simeq 2e$ for $e\leq 0.7$. This implies that  
RS orbits and H orbits merge inside  the inner boundary
of the P orbit domain and  therefore, that collisional orbits are unstable 
in two dimensions. 

In three dimensions, however,  physical collisions occur only when 
the nodes of the asteroid's  orbit cross the orbit of the planet; 
this corresponds to a specific argument of perihelion given by  
$\cos \omega\simeq e$. 
Consequently, the collision singularity of $S$ is generally 
removed and  replaced  
by maxima  whose existence and relative magnitudes depend 
 on  $e$, $I$ and $\omega$. 

This leads to the third major consequence of large $e$ and $I$: the
appearance of {\it compound} and {\it transition} orbits (Fig~2 c,d).
The first correspond to the merger of H or T orbits with RS orbits;
asteroid (3753) Cruithne$^5$ is the first example of a compound H-RS 
orbit in the solar system.  
The second  orbits  correspond to the maxima of $S$; however, they are not 
long-lived because of the secular variations of $e$, $I$ and $\omega$,
in time which modify the magnitudes of the corresponding maxima. 
Transition orbits therefore appear in the orbital evolution only to permit
the passage between different orbit families. 

The secular stability that was found in the three-body problem,$^{6}$ 
together with the presence of asteroid Cruithne 
on an H-RS orbit, suggested the robustness of this 
new dynamics when exposed to planetary perturbations. This motivated 
our search for further examples in the solar system.  

\vspace*{5mm}

\begin{center} {\bf 3. New coorbital asteroids}\\[2mm]\end{center}
We have conducted a search in the 
catalogue of near-Earth objects (NEOs) which is maintained in the Minor 
Planet Center web-site$^{10}$ for objects with semi-major axes closer than 
approximately $\Delta_{\rm P}$  from the semi-major axis of 
a terrestrial planet. 
We carefully avoided objects with poorly determined
orbital elements, choosing to include in our sample only multi-apparition 
asteroids. We stress the need here for further observations 
that refine the orbits of the known near-Earth asteroids.
Our selection criteria yielded a total of five objects, $\sim 1 \%$ of 
the known NEO population, as favorite candidates for transient 
coorbital behavior.
These asteroids are (3753) Cruithne, (3362) Khufu, 1989~VA, 1993~WD and
1994~TF2.
Here, we present results for three of those objects (see Table~1) as
representative of what is observed for the full sample.

The orbits were integrated numerically$^{11}$ with the eight major 
planets (Mercury to Neptune) using a Runge-Kutta-Nystrom 12th order 
scheme with adaptive step-size control.$^{\rm 12}$
This presents an advantage over the scheme used in 
ref~5 due to its ability to model 
close encounters, a vital part of the mechanism 
which gives rise to stable orbit transitions.
The integration time-span was set to be 200,000 years centered on 
the present as a reasonable trade-off between
integrator accuracy, which may deteriorate for longer time-spans,
and the richness of the coorbital dynamics that we expect to observe.

{\it All} the objects examined exhibited transient coorbital behavior. 
Figures 3 and 4 show the orbits'  evolution during the coorbital capture
of (3753) Cruithne and (3362) Khufu by the Earth, and 1989~VA by Venus.
We confirm the existence of the new structures discussed 
in the previous section, such as   
compound  orbits (Cruithne, 1989~VA),  displaced tadpole librations
(Cruithne), retrograde satellites of the Earth (Khufu, Cruithne) and 
the transitions between different coorbital modes. Owing to the
intervening periods of stochastic drift,
mainly in $a_{\rm r}$, we cannot claim that we observe the actual evolution
of the asteroids in the integration time-span. 
What we can say, however, is that each one of those asteroids 
is likely to have been or will be captured in such orbits.  This claim 
is supported by our numerical experiments with slightly modified orbits 
for the asteroids (i.e. clone asteroids).

\vspace*{5mm}

\begin{center}{\bf 4. Discussion}\\[2mm] \end{center} 
Unlike  (3753) Cruithne which is currently on an H-RS orbit, well inside the 
coorbital region of the Earth, 
the remaining 
objects that we examined  are not currently in coorbital libration 
with any planet. 
This fact, however,  did not induce  distinct types of global evolution 
in the sample. For instance, during the evolution, 
Cruithne happened to exit the coorbital 
region;  in contrast,  the other objects could  be captured by 
the Earth or Venus.
This shows that Cruithne's status is less special than previously
thought: it is simply a member of the population
of NEOs that evolve inside the inner solar system 
with a slowly varying semi-major axis.$^{13}$ 
  
This seemingly disappointing conclusion about the true nature 
of Cruithne's evolution can be viewed positively 
in the context of the protection of the Earth from collisions with nearby
asteroids. The evolution of the mentioned asteroids and their clones 
indeed suggests that the Earth and Venus are well protected from the  
class of NEOs that drift (in semi-major axis) towards and 
get close to the coorbital regions of the terrestrial planets.   
A key factor in this protection is the quasi-absence of the chaotic layer
at the edges of the coorbital region, found at small $e$ and $I$,
and responsible for the instability  during the crossing of coorbital 
region.$^{6,14}$  This suggests the presence of undiscovered  
coorbital asteroids of the terrestrial planets, 
that display one of the new coorbital modes. In particular, the connection 
between retrograde satellite orbits and the classic types of motion 
is remarkable because it allows a fraction of the drifting NEOs
to become, temporarily,  retrograde satellites of the terrestrial planets. 
Clear examples of this behavior are the past trapping of Khufu that 
lasted  35,000 years and the future 3,000 year RS phase of Cruithne's 
evolution.  This same connection also offers a new route for the capture
of the known retrograde satellites of the Jovian planets: 
the  dissipative processes  of 
early planetary formation such as the planet's growth$^{15}$  or 
the drag due to  the circumplanetary nebula,$^{16}$  can produce  
permanent capture by reducing the Jacobi constant of a Trojan 
asteroid that undergoes 
a transient RS motion. This results in a distant retrograde satellite with 
large planetocentric eccentricity and inclination, which is  reminiscent 
of the orbits of the known retrograde satellites of the 
Jovian planets.$^{17}$

The evolution of our sample shows that the coorbital asteroids of Earth
and Venus  with large $e$ and $I$  are generally  expelled from 
the coorbital regions because of  the  close encounters with the other 
planets. In that respect, Mercury is a better candidate for finding 
permanent coorbitals because of its distant location from Venus. 
The presence of a specific population of minor bodies close to Mercury  
has already been proposed in the context of the vulcanoid hypothesis 
for the chronology of the  geological evolution the planet.$^{18}$  However,
the dynamical constraints of the model assumed  that the vulcanoids
had small eccentricities and inclinations and the recent 
observational searches$^{18}$ produced negative results. 
In view of (1) the large eccentricity 
of Mercury ($e=0.2$), (2) the action of nearby secular resonances 
and (3) the stability of the structures we found, 
the putative coorbital vulcanoids are likely to have stable large 
eccentricity and inclination orbits. The evidence of such objects 
would argue for the early presence and slow depletion of a 
larger reservoir of bodies that contributed  significantly to the cratering 
history of Mercury.

\begin{small}

\end{small}

\newpage
\vspace*{3mm}
\noindent {\bf Figure captions}
\begin{enumerate}
\item Representative zero velocity curves for three values of the 
Jacobi constant and  the mass ratio $m_{\rm P}/(m_{\rm P}+m_{\rm S})=0.01$.  
The locations of the Lagrangian equilibrium points $L_1$--$L_5$ are 
indicated by small open circles.  The dashed line denotes a circle 
of radius equal to the 
planet's semi-major axis. The letters T (tadpole), 
H (horseshoe) and P (passing) denote the type of orbit associated with the 
curves.  The regions enclosed by each curve (shaded) are excluded from the 
spatial motion of the asteroid that has the corresponding value of the 
Jacobi constant. 
Note that the curve between the $L_1$ and $L_2$ points permits 
satellite orbits around the planet inside the Hill sphere of radius 
$\Delta_{\rm P}$ (Eq~1).

\item Types of coorbital motion. $S$ is viewed as an effective 
potential well and the possible guiding centers are given by the levels 
$3(m_{\rm P}+m_{\rm S})C/8m_{\rm P}$ (light lines). The dashed levels 
correspond to transition orbits. T, H, RS and P denote respectively 
tadpole, horseshoe and retrograde satellite and passing orbits. 
The hyphenated designations denote compound orbits. $L_1$ and $L_2$ 
are located at $\lambda_{\rm r}=0^\circ$ and $L_3$ at $\pm 180^\circ$. 
The bold ticks denote the locations of $L_4$ and $L_5$.
The comparison of {\bf a}  (circular 2D orbits) and {\bf b}  
(eccentric 2D orbits, $e=0.3$) shows the appearance of retrograde satellite 
orbits outside the Hill sphere and the displacement of $L_4$ and $L_5$ form 
the equilateral configuration  $\lambda_{\rm r}=\pm60^\circ$.
In 3D, ({\bf c}:  $e=0.3$, $I=20^\circ$, $\omega=60^\circ$;  
{\bf d}: $e=0.5$, $I=30^\circ$, $\omega=0^\circ$), orbit transition 
and compound orbits become possible. 
Note the asymmetry in the location and the maxima of  $L_4$ and $L_5$.
Because  $S(\lambda_{\rm r},e,I,-\omega)=S(-\lambda_{\rm r},e,I,\omega)$,
the coorbital motion also includes orbits that are symmetric to those in 
{\bf c}  with respect to $\lambda_{\rm r}=0$.

\item Coorbital capture of near-Earth asteroids. The evolution of the  
relative semi-major axis, $a_{\rm r}$, and  
the difference in mean longitudes between the object and
the relevant planet, $\lambda_{\rm r}$, are shown as functions of time.
The plots correspond to (3753) Cruithne ({\bf a, b}) and  (3362) Khufu 
({\bf c, d}) coorbiting with the Earth, and 1989~VA ({\bf e, f})
coorbiting with Venus. The asteroids are seen to transit between 
many types of coorbital motion (Fig~2) for time-scales ranging 
from thousands to tens of thousands of years.
Note that in all three cases, the transitions  occur without the large jumps in
semi-major axis which are indicative of unstable close approaches. 

\item Coorbital librations of the guiding center. 
The trajectories observed in Fig~4 are shown 
in the ($\lambda_{r},a_{r}$) plane. {\bf a}: (3753) Cruithne, {\bf b}: 
(3362) Khufu and {\bf c}: 1989~VA. Note that the proximity of Khufu's 
guiding center ($\sim 1^\circ$) to the Earth  (of radius $\simeq 9^{\prime
\prime}$) 
does not result in a 
near collision because the full motion of the asteroid includes the 
fast gyration that was averaged out and whose smallest amplitude 
is $\sim a_{\rm A}e$. This  means that 
the asteroid is  $\sim 0.5$ {\sc au} away from the    
Earth. The temporary proximity to $\lambda_{\rm r}=0^\circ$ implies that
the trajectory associated with the fast gyration is centered on the Earth 
and  does not precess secularly. 
\end{enumerate}

\begin{table}[htb]
\caption{Osculating elements for three coorbital candidates at JD 2451000.5 
as given by the JPL HORIZONS ephemeris generation system.$^{19}$
$a_{\rm A}$, $e$, $I$, $\omega$ and $\Omega$ denote respectively the
semi-major axis, the eccentricity, the inclination, the  
argument of perihelion and the longitude of the ascending node.
$a_{\rm r}a_{\rm P}$ is the difference between the semi-major axes 
of the coorbital 
planet and the asteroid; it is given in terms of 
the width of the planet's coorbital region $\Delta_{\rm P}$.
OCC$^{19}$ (Orbit Condition Code) provides a rating of the quality of the
computed orbit based on the available observations. In this scale 0 is the 
best and 7 the worst with 1 the typical value for numbered asteroids.}
\label{objects}
\begin{tabular}{lrrrrrrr}
\noalign{\smallskip}
\hline
Object & $a_{\rm A}$ ({\sc au}) & $e$ & $I\ (^{\circ})$ & $\omega\ (^{\circ})$ 
& $\Omega\ (^{\circ})$ & $a_{\rm r}a_{\rm P}$ & OCC \\ \hline
(3362) Khufu    &  0.9895  & 0.469  & 9.9 &  54.8  & 152.6  & $1.05\, 
\Delta_{\rm Earth}$ & 1 \\
(3753) Cruithne &  0.9978  & 0.515  & 19.8  & 43.7  & 126.4  & $0.22\, 
\Delta_{\rm Earth}$ & 1 \\
1989~VA         &  0.7287  & 0.595  & 28.8  & 2.8  & 225.7  & $0.57\, 
\Delta_{\rm Venus}$ & 3 \\\hline
\end{tabular}
\end{table}

\end{document}